\documentstyle[11pt,aaspp4]{article}
\slugcomment{Submitted to the Astrophysical Journal}
\input psfig.sty

\def\grb{GRB\thinspace{970228}}
\def\ts{\thinspace}

\begin{document}

\title{No Radio Afterglow from the Gamma-Ray Burst of February 28, 1997}

\author{D. A. Frail\altaffilmark{1}, S. R. Kulkarni\altaffilmark{2}, 
D. S. Shepherd\altaffilmark{2}, 
E. Waxman\altaffilmark{3}} 
 
\altaffiltext{1}{National Radio Astronomy Observatory, Socorro,
           NM, 87801, USA} 

\altaffiltext{2}{Division of Physics,
Mathematics and Astronomy 105-24, Caltech, Pasadena CA 91125, USA}

\altaffiltext{3}{Institute for Advanced Study, Princeton, NJ 08540, USA}

\begin{abstract}
  We present radio observations of the gamma-ray burster \grb\ made
  with the Very Large Array (VLA) and the Owens Valley Radio
  Observatory (OVRO) spanning a range of postburst timescales from one
  to 300 days. A search for a time-variable radio source was conducted
  covering an area which included a fading X-ray source and an optical
  transient, both of which are thought to be the long wavelength
  counterparts to the gamma-ray burst. At the position of the optical
  transient sensitive limits between 10 $\mu$Jy and 1 mJy can be
  placed on the absence of a radio counterpart to \grb\ between 1.4
  and 240 GHz. We apply a simple formulation of a fireball model which
  has been used with some success to reproduce the behavior of the
  optical and X-ray light curves.  Using this model we conclude that
  the radio non-detections are consistent with the peak flux density
  of the afterglow lying between 20-40 $\mu$Jy and it requires that
  the optical flux peaked between 4 and 16 hours after the burst.

\end{abstract}


\vfill\eject
\section{Introduction}

The gamma-ray burst of 28 February 1997 was a turning point in our
understanding of these enigmatic objects, resulting in the first-ever
discovery of X-ray and optical counterparts to a burst.  Within the
original 3-arcminute localization provided by the Wide Field Cameras
(WFC) on board the BeppoSAX satellite, a previously unknown X-ray
source 1SAX\ts{J0501.7+1146} was detected by the Narrow Field
Instruments (NFI). Eight hours after the burst the flux of
1SAX\ts{J0501.7+1146} was 2.8$\times{10}^{-12}$ erg cm$^{-2}$ s$^{-1}$
(2-10 keV), but three days later its flux had dropped by a factor of
20 (Costa et al. 1997). A comparison of V- and I-band images taken on
28 February and 8 March revealed an optical transient within a reduced
error box, defined by the intersection of the WFC circle, the
$\pm$50\arcsec\ circle of the NFI, and the Interplanetary Network
(IPN) annulus (van Paradijs et al. 1997, Hurley et al. 1997).

Predictions of long-lived afterglows from gamma-ray bursts at X-ray,
optical and radio wavelengths have been made for some time (e.g.
Paczy\'nski \& Rhoads 1993, M\'esz\'aros, Rees, \& Papathanassiou
1994, Katz 1994, M\'esz\'aros \& Rees 1997). In particular, a
power-law decay in the observed long-wavelength flux with time is a
generic consequence of a class of models known as ``fireballs''. The
gamma-rays are thought to be produced when the relativistically
expanding blast wave (aka fireball) is slowed down by the ambient gas
or is self-shocked by its own ejecta. The fireball accelerates
particles in a shock which then radiate via the synchrotron process
(M\'esz\'aros et al. 1994, Waxman 1997a, 1997b, Sari, Piran, \&
Narayan 1998). As the fireball expands it cools, shifting the peak in
the spectrum to lower energies and resulting in delayed emission at
longer wavelengths.

The optical and X-ray decay from \grb, as well as that from several
other subsequent GRBs, is consistent at least to first order with one
of the simpler formulations of these models (M\'esz\'aros \& Rees
1997). Costa et al. (1997) fit a t$^\delta$ decay to the X-ray data
with $\delta\simeq-1.33\pm{0.12}$, while global fits to the X-ray,
optical and infrared data give $\delta=-1.2$ (Wijers, Rees and
M\'esz\'aros 1998) and $\delta=-1.09\pm0.23$ (Reichart 1997). More
recent optical fits, aided by a long time-baseline, yield
$\delta=-1.12\pm0.08$ (Garcia et al. 1997), $\delta=-1.21\pm0.02$
(Masetti et al. (1997), and $\delta=-1.10\pm0.04$ (Galama et
al. 1998). The character of this decay agrees well with a fireball
produced by a one-time impulsive injection of energy in which only the
forward blast wave efficiently accelerates particles (i.e. the
adiabatic piston model). The adiabatic model predicts a simple
relation between the slope of the temporal decay $\delta$ and the flux
spectrum $\beta$=${2\over3}\delta$. Observationally the spectrum is
not well determined, with $\beta=-0.7\pm0.1$ optically (van Paradijs
et al. 1997) and X-ray values of $\beta\simeq-0.9$ with large scatter
(Frontera et al. 1998).  The slope of the particle spectrum $p$ (where
$\beta=(p-1)/2)$ inferred from these values of temporal decay is
$p\sim-2.6$, not an unreasonable value for a relativistic shock
(Blandford and Eichler 1987).

Given the early success of the fireball model in predicting the gross
properties of the optical and X-ray behavior of \grb, it is reasonable
to look for delayed radio emission from this burst. This is
particularly relevant in the light of the discovery of the radio
afterglow from GRB\thinspace{970508} (Frail et al. 1997a) which
exhibited temporal and spectral behavior consistent with this model
(Waxman, Kulkarni \& Frail 1997).  The properties of the fireball for
\grb\ are well constrained by the optical and X-ray data.  Thus the
presence or absence of radio emission at late times has a bearing on
the validity of this model. With this in mind, we began a radio search
centered on the optical transient detected by van Paradijs et
al. (1997). This {\it Letter} is a summary of our monitoring program
for the first 300 days.

\section{Observations and Results}

Radio observations of a field centered on \grb\ began with the Very
Large Array (VLA) on 1997 March 1.03 UT, 22 hours following the burst.
Initially the VLA observations were made at a frequency of 1.43 GHz
($\lambda$=21 cm), where the field of view is large enough to
accommodate the preliminary 10\arcmin\ radius error circles from
BeppoSAX.  As the position of \grb\ became better known we increased
the observing frequency first to 4.63 GHz ($\lambda$=6.5 cm) and then
to 8.46 GHz ($\lambda$=3.5 cm), beginning three days after the first
VLA observation. On 1997 March 16.09 UT a full synthesis single track
was made with the millimeter interferometer at the Owens Valley Radio
Observatory (OVRO) at 92.7 GHz ($\lambda$=3.2 mm) and 242.2 GHz
($\lambda$=1.2 mm). See Shepherd et al. (1998) for more details on the
OVRO observation.

The results of our VLA and OVRO observing campaign of \grb, spanning
postburst timescales from one to 300 days, are summarized in Table 1
and are plotted in Figure 1. The dates of each observation are given
followed by the number of elapsed days, the observing frequency, the
synthesized beam size and the rms noise on each image. The last column
in Table 1 quantifies our principal scientific result, that between
one and 300 days after the initial gamma-ray event no radio emission
was detected from \grb. One radio source was detected in the
$\pm$50\arcsec\ radius of the NFI (Frail et al. 1997b), but it lies
outside the reduced IPN/WFC/NFI error box and has spectral and
temporal characteristics that are consistent with it being an
unrelated background extragalactic object.  At the location of the van
Paradijs et al. (1997) optical transient sensitive 1$\sigma$ limits
between 10$\mu$Jy and 1 mJy can be placed on the {\it absence} of a
radio counterpart to \grb\ between 1 and 250 GHz. The limits at 8.46
GHz are the most severe, with typical 2$\sigma$ limits of 100 $\mu$Jy.
Adding the 8.46 GHz data from all 14 epochs together produces a
2$\sigma$ upper limit of 9 $\mu$Jy. The Westerbork Radio Synthesis
Telescope (WSRT) and the Berkeley-Illinois-Maryland Array (BIMA) also
observed \grb\ at 5 GHz and 86.4 GHz, respectively, producing upper
limits a factor of two or more above those in Table 1 (Groot et al.
1997, Smith et al. 1997).

\section{Discussion}

\subsection{The Adiabatic Piston Model}

In this section we will briefly review the spectral and temporal
behavior for the impulsive, forward shock dominated fireball model. It
is treated in far more detail by M\'esz\'aros \& Rees (1997), Waxman
(1997a, 1997b) and Sari et al. (1997). We will not consider radiative
models nor treat the case of rapid cooling (Vietri 1997b).  The
optical light curve of \grb\ has maintained its power law slope with
no break even six months after the initial detection (e.g. Garcia et
al. 1997), suggesting that the adiabatic model is an appropriate
description of this system. Moreover, the absence of such a break
suggests that the fireball is not far from being spherically symmetric
(i.e. that it is not a jet).

Initially the fireball ejecta have a bulk Lorentz factor
$\Gamma$=10$^2$-10$^3$ which falls as t$^{-3/8}$, where t is elapsed
time in the observer reference frame. The blast wave ahead of the
ejecta accelerates a power-law distribution of electrons with a
minimum Lorentz factor $\gamma_{m}$ and a power law index of p. The
emergent spectrum in the co-moving frame has a synchrotron break at a
frequency $\nu_m\propto\Gamma$B$^\prime\gamma_{m}^2$ (where B$^\prime$
is the co-moving magnetic field). Most of the radiation is emitted
near the synchrotron break frequency. Above $\nu_m$ the spectral slope
$\beta$=(1$-p$)/2 is set by the shock processes. Below $\nu_m$ the
spectral slope $\alpha$ is predicted to be 1/3 from synchrotron theory
(Pacholczyk 1970, Katz 1994) but it could be flatter if there are a
range of physical conditions (i.e. B$^\prime$, $\Gamma$) in the shock.

The delayed emission at X-ray, optical and radio wavelengths
originates from the deceleration of the fireball, which produces a
shift of $\nu_m$ to longer wavelengths. The X-ray and optical flux
will reach a maximum F$_m$ when $\nu_m$ moves into their respective
wavelength bands with a time-dependence $\nu_m\propto$t$^{-3/2}$.  The
flux when $\nu<\nu_m$ rises as t$^{3\alpha/2}$ to a peak of F$_m$,
thereafter decaying as t$^{3\beta/2}$ when $\nu>\nu_m$. There are two
significant points to note here, first the frequency $\nu_m$ is
independent of the fraction of fireball energy initially carried by
rest mass and of the density of the medium into which it is expanding,
and second, the peak flux density F$_m$ is the same at all
frequencies. We further note, that as a general result it is easy to
show that the radio emission increases until $\Gamma\sim$a few, when
the shock is only weakly relativistic. Since all the terms in $\nu_m$
depend linearly on $\Gamma$, $\nu_m\propto\Gamma^4$. At the time of
the gamma-ray burst $\Gamma_\circ\sim{300}$ and $\nu_m\sim10^{20}$ Hz,
so scaling to the radio band ($\nu_m\sim10^{10}$ Hz) we obtain
$\Gamma\sim{1}$.

The above behavior assumes that the emission emerges from an optically
thin fireball. Synchrotron self absorption was predicted (Katz 1994,
Katz \& Piran 1997) to be important at low radio frequencies during
the first few weeks of the fireball's existence. For
GRB\thinspace{970508} Shepherd et al. (1998) observed a low frequency
turnover radio spectrum of the afterglow during the first week,
consistent with synchrotron self absorption with an optical depth of
one near 1.6 GHz. Frail et al. (1997a) observed a gradual rise in the
1.43 GHz flux at later times consistent with the fireball becoming
optically thin at this frequency. The spectral slope in the
self-absorbed part of the spectrum is $\nu^2$ when the shock is
relativistic but becomes the familiar $\nu^{5/2}$ when $\Gamma\sim{1}$
(Katz 1994, Waxman 1997b, Katz \& Piran 1997). Early estimates
erroneously used a synchrotron self-absorption frequency $\nu_A$ of
$\sim$10$^{13}$ Hz (e.g. M\'esz\'aros \& Rees 1997) but later
corrections have shifted $\nu_A$ to between 1-5 GHz (Waxman 1997b). In
this paper it is pointed out that the addition a low energy population
of electrons below $\gamma_{m}$ can increase the synchrotron opacity
without increasing F$_m$, provided that these electrons are small in
number compared to the high energy population. While this increases
$\nu_A$ somewhat,the fact remains that beyond 1-2 weeks after the
burst the fireball emission should be optically thin at 8.46 GHz where
most of our observations have been made.

\subsection{The radio data and constraints on the fireball model}

To first order the model described in \S{3.1} gives a fairly good
description of the behavior of the X-ray and optical afterglow from
\grb. We will now use it to predict the late-time behavior of the
afterglow at radio wavelengths. A simple-minded extrapolation of the
spectral slope $\alpha$, measured at the time when the optical and
X-ray transients were detected, would predict flux densities in the
centimeter range of 10-100 mJy. Such flux densities were clearly not
seen so we can infer that at this time the radio band $\nu_r$ was
below $\nu_m$. Frequencies below 5 GHz may have been initially
strongly self absorbed (i.e. $\nu_r<\nu_{ab}<\nu_m$) but this would
not be the case for the majority of our data at 8.46 GHz nor our OVRO
data at millimeter wavelengths (Waxman 1997b).

The radio emission peaks at F$_m$ at a time t$_r$, delayed from the
optical peak given by t$_r$=t$_o(\nu_r/\nu_o)^{-2/3}$. For
$\nu_r$=8.46 GHz t$_r\simeq{1400}$\ t$_o$ for $\nu_r$=92.7 GHz and
t$_r\simeq{300}$\ t$_o$. Neither F$_m$ nor t$_o$ are well constrained
by the X-ray and optical afterglows since a distinct turnover was not
observed. One approach used by Wijers et al. (1997) is to use the
properties of the gamma-ray burst to constrain the values of F$_m$ and
t$_o$ for the fireball. There is no a priori reason to assume that the
GRB and the fireball are directly related in this way. In fact, it is
currently held that the two are physically decoupled from one other
(e.g. M\'esz\'aros 1997).

Nevertheless, if we adopt t$_o\simeq670$ s for R band used by Wijers
et al. we find that the predicted radio emission from \grb\ would peak
at $\sim10^{6}$ s, thereafter decaying in the same manner as the
optical and X-ray.  At 8.46 GHz we have sensitive upper limits for the
absence of radio emission at the location of the optical transient,
spanning approximately two orders of magnitude in time (10$^5$-10$^7$
s).  For the value of F$_m\simeq{5000}$ $\mu$Jy and the t$^{-1.2}$
decay fit for $\nu>\nu_m$ from Wijers et al. (1997), the VLA would
have been expected to detect a bright radio source during its entire
monitoring interval ($>$100 $\mu$Jy). Our millimeter observations at
OVRO are less constraining.  At the time of these observations, nearly
two weeks after the burst, the millimeter emission from \grb\ is
expected to have decayed below detectablity and thus it cannot provide
any useful constraints on the fireball model. Wijers et al. (1997)
assume $\alpha=0$ for \grb, so the flux is constant and equal to F$_m$
until $\nu=\nu_m$. A more conservative approach is to set $\alpha$ to
the asymptotic value of $1/3$, so that the flux rises as t$^{1/2}$.
This is still at odds with our radio data and we conclude that the
estimates of F$_m$ and t$_o$ used by Wijers et al. are incorrect.

The early optical light curve for \grb\ provides additional
constraints on F$_m$ and t$_o$. In the adiabatic model the optical
flux at any time t provided $\nu_o>\nu_m$ (i.e. the decay phase) is
given by F$_o$=F$_m$($\nu_o/\nu_m)^\beta$, or
F$_o$=F$_m$(t/t$_o)^{3/2\beta}$.  At the same time provided
$\nu_r<\nu_m$ (i.e. rising phase) then
F$_r$=F$_m$($\nu_r/\nu_m)^\alpha$ or F$_r\propto$t$^{1/2}$ for
$\alpha=1/3$ (Waxman 1997a, 1997b). We can solve these two equations
to relate the rising radio flux density at any time in terms of the
decaying optical flux at the same time.
$${\rm F}_r\ =\ {\rm F_o}\ (\nu_r/\nu_o)^{\alpha}\
{\rm(t/t}_o)^{3/2(\alpha-\beta)}$$
\noindent{As noted earlier the radio flux at 8.46 GHz is expected to
rise to a maximum F$_m$ at a time t$_r\simeq{1400}$\ t$_o$ thereafter
it would decay identically to the optical and X-ray. Small values of
t$_o$ give rise to large values of F$_m$.

A compilation of flux measurements for the optical transient are given
by Galama et al. (1997) and updated in Galama et al (1998). The
earliest measurement is by Pedichini et al. (1997) on February 28.81
UT which was translates to an R-band magnitude of 20.5$\pm$0.5 (Galama
et al. 1997). A similar but more accurate value of 20.9$\pm$0.14 was
measured on February 28.99 UT by van Paradijs et al. (1997).
Converting these R-band magnitudes to flux densities we derive values
of 18$^{+11}_{-7}$ $\mu$Jy and 12$\pm$2 $\mu$Jy, respectively. No
clear signature of a turnover has been seen in the optical data and in
fact the power law decay of \grb\ has maintained its character with
small deviations beginning 16 hrs after the burst to the present time
(Galama et al. 1998). Thus the peak flux density must have been prior
to this time, for which we derive F$_m\geq$18 $\mu$Jy and
t$_o\leq6\times{10}^4$ s.

The radio upper limits in Table 1 can constrain F$_m$ and t$_o$
further still. Using the fit to the optical light curve from Galama et
al. (1998) and the above equation, we plot a number of hypothetical
radio light curves in figure 1 for t$_o$=6$\times10^2$ s,
6$\times10^3$ s, and 6$\times10^4$ s. The absence of readily
detectable radio emission 300 days after the original event requires
t$_o>$2.8$\times10^4$ s and F$_m<40$ $\mu$Jy. Similarly, an upper
limit on t$_o$ is set by the date of the earliest detection of optical
emission from \grb\ of 18 $\mu$Jy at 6$\times10^4$ s by Pedichini et
al. (1997). If the adiabatic fireball model is to remain a valid
description of the afterglow emission from \grb\ then it predicts that
the radio afterglow will peak around 40 to 18 $\mu$Jy between
2$\times10^7$ s and 8$\times10^7$ s. Castander \& Lamb (1998) have
argued in favor of visual extinction A$_V$=1.3 towards \grb, a value
some two to three times larger than used in most papers to date. This
has the effect of raising F$_m$, making the radio non-detections even
more problematic. The radio emission may be weaker than these
predictions when the decelerating fireball eventually becomes
sub-relativistic (Wijers et al. 1997). However, such a transition
should first be seen in the optical light curve as a break in the
power law slope and no such break has been seen thus far.

\section{Conclusions}

The detection of an optical transient, thought to be the afterglow
from \grb, has made it possible to perform a search for time-variable
radio emission at the optical position. Delayed emission at longer
wavelengths (X-ray, optical and radio) is a generic prediction of all
fireball models. VLA and OVRO observations have been made that span a
range of postburst timescales from one to 300 days putting limits on
the absence of a radio counterpart to \grb\ between 10 $\mu$Jy and 1
mJy.  Applying a simple version of the fireball model which has been
used successfully to fit the temporal behavior of the decaying X-ray
and optical emission from \grb\ suggests that the radio afterglow has
yet to peak. If this is correct then continued deep imaging in the
coming months offers the possibility for the detection of a weak but
increasing radio signal. Detecting this emission would be a powerful
verification of the fireball model as described by M\'esz\'aros \&
Rees (1997).

Indeed, as the recent detection of GRB\thinspace{970508} has taught
us, radio afterglows yield unique GRB diagnostics that are not
obtainable by any other means (Frail et al. 1997b).  Unlike optical or
X-ray wavelengths one is presented with the possibility of following
the {\it full} evolution of the fireball emission through all its
different stages; first while it is optically thick, then as it slowly
rises to a peak flux density and thereafter decays, making a
transition from an ultra-relativistic to sub-relativistic shock.
Furthermore, both the scintillation of the radio source (Goodman 1997)
and its flux density, when it is synchrotron self-absorbed (Katz
1994), allow a determination of the size and expansion of the
fireball.  If the radio afterglow is going to be detected from \grb\ 
and others like it, then continued long-term monitoring is going to be
required at the microJansky level.

\acknowledgments

The VLA is a facility of the National Science Foundation operated
under cooperative agreement by Associated Universities, Inc. SRK's
research is supported by the National Science Foundation and NASA.
DAF thanks Ralph Wijers for useful discussions.

\centerline{\hbox{\psfig{file=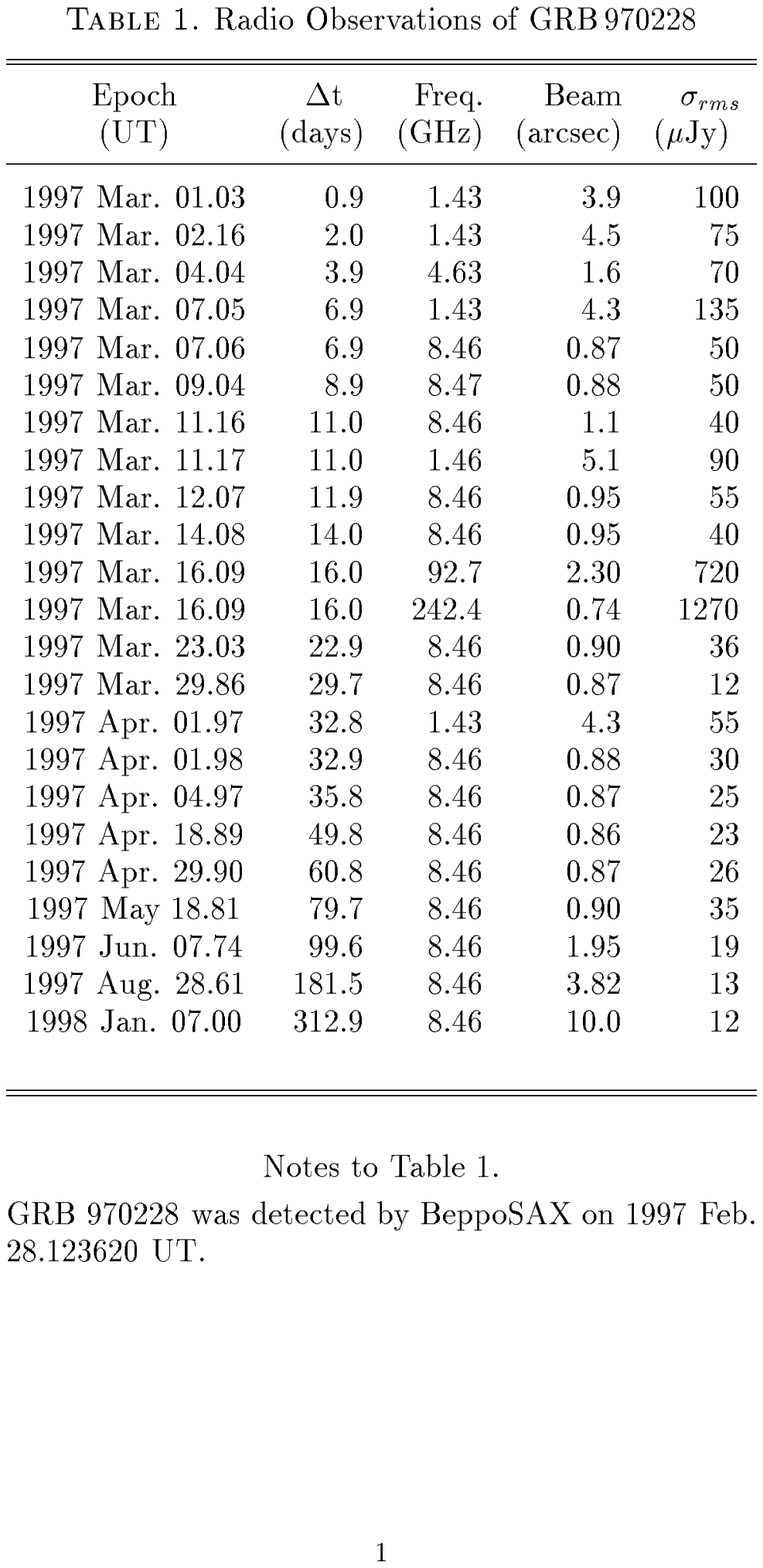,width=8.5in,clip=0}}}
\medskip

\clearpage
\centerline{\hbox{\psfig{file=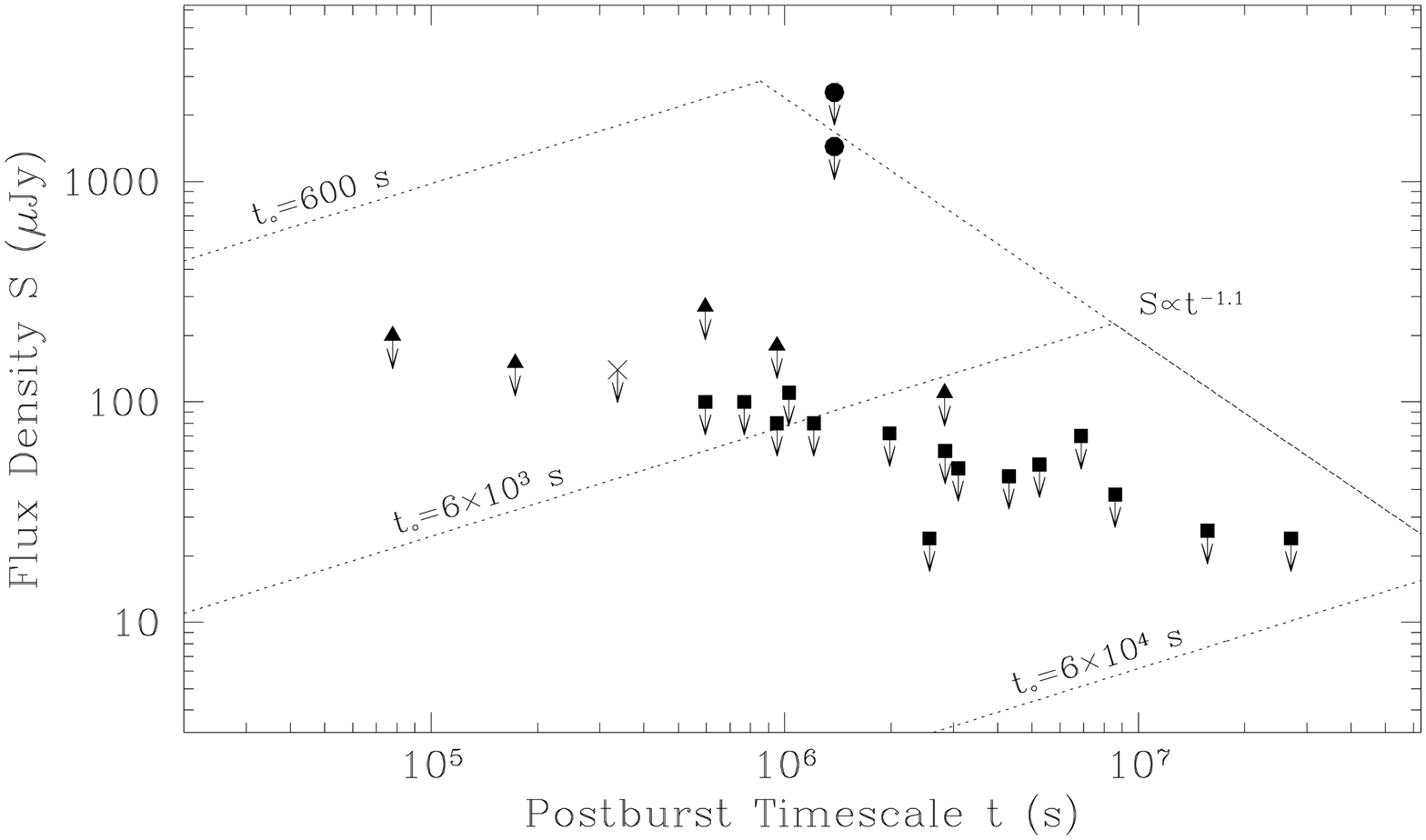,width=7.0in,clip=0,angle=270}}}
\medskip
\noindent {\bf Figure 1.--} Upper limits on the flux density of a
radio counterpart to the optical transient associated with \grb. VLA
points are given by the filled triangles (1.4 GHz), a cross (4.6 GHz)
and the filled squares (8.5 GHz). OVRO points are the filled circles
at 240 GHz (top) and 90 GHz (bottom). The dashed lines are the
expected radio light curves at 8.46 GHz for the adiabatic piston model
using the observed optical decay (Galama et al. 1998) scaled to the
radio and assuming several values for the timescale of the onset of
the decay phase (see text for more details).


\clearpage
 
\begin{references}

\reference {be87} Blandford, R., \& Eichler, D. 1987, Phys. Rep., 154, 1

\reference {cl98} Castander, F. J., \& Lamb, D. Q. 1998, in 4th
Huntsville GRB Symposium, eds. C. Meegan, R. Preece and T. Koshut, (New
York:AIP), in press

\reference {cos97} Costa, E. et al. 1997, Nature, 387, 783

\reference {fra97a} Frail, D. A. et al. 1997a International Astr. Union
Circular, No. 6576

\reference {fra97b} Frail, D. A., Kulkarni, S. R., Nicastro, L.,
Feroci, M., \& Taylor, G. B. 1997b, Nature, 389, 263

\reference {ron97} Frontera, F. et al. 1998, \apj, 493, L67

\reference {gal97} Galama, T. et al. 1997, Nature, 387, 479

\reference {gal97} Galama, T. et al. 1998, in 4th Huntsville GRB
Symposium, eds. C. Meegan, R. Preece and T. Koshut, (New York:AIP), in
press

\reference {gar97} Garcia, M. R. et al. 1997, \apj\ submitted (LANL
Preprint astro-ph/9710346)

\reference {goo97} Goodman, J. 1997, New Ast, 2, 449

\reference {gro97} Groot, P. J. 1997, International Astr. Union
Circular, No. 6584

\reference {hur97} Hurley, K. et al. 1997, \apj, 485, L1

\reference {kat94} Katz, J. J. 1994, \apj, 422, 248

\reference {kp97} Katz, J. J., \& Piran, T. 1997, \apj, 490, 772

\reference {mas97} Masetti, N., Bartolini, C., Guarnieri, A., \&
Piccioni, A. 1997, LANL Preprint astro-ph/9711260

\reference {m97} M\'esz\'aros, P. 1998, in 4th Huntsville GRB
Symposium, eds. C. Meegan, R. Preece and T. Koshut, (New York:AIP), in
press

\reference {RM97} M\'esz\'aros, P., Rees, M. J., \& Papathanassiou, H.
1994, \apj, 432, 181

\reference {RM97} M\'esz\'aros, P. \& Rees, M. J. 1997, \apj, 476, 232

\reference {p70} Pacholczyk, A. G. 1970, Radio Astrophysics (San Francisco:
W. H. Freeman)

\reference {PR94} Paczy\'nski, B. \& Rhoads, J. E. 1993, \apj, 418, L5

\reference {ped97} Pedichini, F. et al. 1997, A\&A, 327, L36

\reference {R97} Reichart, D. E. 1997, \apj, 485, L57

\reference{spn97} Sari, R., Piran, T., \& Narayan, R. 1998, \apj,
submitted (LANL Preprint astro-ph/9712005)

\reference{She97} Shepherd, D. S., Frail, D. A., Kulkarni, S. R., \&
Metzger, M. R. 1998, \apj, 497, 859

\reference{smi97} Smith, I. A., Gruendl, R. A., Liang, E. P., \& Lo,
K. Y. 1997, \apj, 487, L5

\reference {van97} van Paradijs et al. 1997, Nature, 386, 686

\reference {vie97} Vietri,  M. 1997a, \apj, 478, L9

\reference {vie97} Vietri,  M. 1997b, \apj, 488, L105

\reference {wax97a} Waxman, E. 1997a, \apj, 485, L5

\reference {wax97b} Waxman, E. 1997b, \apj, 489, L33

\reference {wax97b} Waxman, E., Kulkarni, S. R., \& Frail, D. A. 1997,
\apj, 497, 288

\reference {wrm97} Wijers, R. A. M. J.,  Rees, M. J., \& 
M\'esz\'aros, P. 1997, \mnras, 288, L51

\end{references}
\end{document}